\documentclass{pasj00}
\draft

\begin{document}

\SetRunningHead{Hayasaki et al.}{Mass Function of Binary Massive Black Holes in Active Galactic Nuclei}
\Received{2010/1/18}

\title{Mass Function of Binary Massive Black Holes in Active Galactic Nuclei}


\author{Kimitake \textsc{Hayasaki}}
\altaffiltext{}{
Department of Physics, Graduate School of Science, Hokkaido University
Kitaku, Sapporo 060-0810, Japan
}
\email{hayasakin@astro1.sci.hokudai.ac.jp}
\author{Yoshihiro \textsc{Ueda}}
\affil{Department Astronomy, Kyoto University 
Oiwake-cho, Kitashirakawa, Sakyo-ku, Kyoto 606-8502 }
\author{Naoki \textsc{Isobe}}
\affil{Department Astronomy, Kyoto University 
Oiwake-cho, Kitashirakawa, Sakyo-ku, Kyoto 606-8502 }

\def\red#1{\textcolor{red}{{#1}}}
\def\blue#1{\textcolor{blue}{{#1}}}
\def\green#1{\textcolor{green}{{#1}}}

%


\maketitle

\begin{abstract}
If the activity of active galactic nuclei (AGNs) is predominantly 
induced by major galaxy mergers, then a significant fraction 
of AGNs should harbor binary massive black holes in their centers.
We study the mass function of binary massive black holes 
in nearby AGNs based on the observed AGN black-hole mass function and 
theory of evolution of binary massive black holes 
interacting with a massive circumbinary disk 
in the framework of coevolution of massive black holes and their host galaxies.
The circumbinary disk is assumed to be steady, axisymmetric, geometrically thin, self-regulated, self-gravitating but non-fragmenting
with a fraction of Eddington accretion rate, 
which is typically one tenth of Eddington value.
The timescale of orbital decay is {then} estimated as 
$\sim10^8\rm{yr}$ for equal mass black-hole, 
being independent of the black hole mass, semi-major axis, and viscosity parameter
but dependent on the black-hole mass ratio, Eddington ratio, and mass-to-energy conversion efficiency. 
This makes it possible for any binary massive black holes to merge within a Hubble time by the binary-disk interaction.
We find that $(1.8\pm0.6\%)$ for the equal mass ratio and $(1.6\pm0.4\%)$ for the one-tenth mass ratio 
of the total number of nearby AGNs have close binary massive black holes 
with orbital period less than ten years in their centers, 
detectable with on-going highly sensitive X-ray monitors such as 
{\it Monitor of All-sky X-ray Image} and/or {\it Swift}/Burst Alert Telescope. 
Assuming that all binary massive black holes have the equal mass ratio, 
about $20\%$ of AGNs with black hole masses of $10^{6.5-7}M_{\odot}$ 
has the close binaries 
and thus provides the best chance to detect them.
\end{abstract}



\section{Introduction}
\label{sec:intro}

Most galaxies are thought to have massive black holes at their centers \citep{kr95}. 
Massive black holes play an important role 
not only in the activities of active galactic nuclei (AGNs) and quasars 
but also in the formation and evolution of galaxies\citep{mag98,fm00,geb00}. 
Galaxy merger leads to the mass inflow to the central region
by tidal interactions and then a nucleus of the merged galaxy is activated 
and black hole grows by gas accretion \citep{yutre02}.
At some step, the outflow from the central black hole sweeps away 
the surrounding gas and quenches the star formation 
and black hole growth. 
This also produces an observed correlation between
the black hole mass and velocity dispersion 
of individual galaxies \citep{dim05}.

During a sequence of processes, 
binary massive  black holes with a subparsec-scale 
separation are inevitably formed before two black holes merge by emitting 
gravitational radiation. Recent hydrodynamic simulations showed the rapid binary black hole formation
in the parsec scale within several $\rm{Gyrs}$ by the interaction between the black holes and 
the surrounding stars and gas in gas-rich galaxy merger \citep{do07,may07}.
Even if there are transiently triple massive black holes in a galactic nucleus, 
the system finally settles down to the formation of binary massive black holes
by merging of two black hole or by ejecting one black hole from the system 
via a gravitational slingshot \citep{iwa06}.

In coalescing process of two massive black hole,
there has been the so-called final parsec problem: 
it is still unknown how binary massive black holes evolve
 after its semi-major axis reached to the subparsec scale 
where the dynamical friction with the neighboring stars is no longer effective.
Many authors have tackled the final parsec problem in the context of the interaction
between the black holes and the stars, but there has been still extensive discussions
\citep{bege80, makino97, q97, milo03, se07, matsu07, matsui09}.
There is other possible way to extract the energy and 
the angular momentum from binary massive black holes 
by the interaction between the black holes and the gas surrounding them.
This kind of the binary-disk interaction  
could also be the candidate to resolve the final parsec problem 
\citep{iv99, go00, armi1, armi2, es05, haya09, cu09, ha09} in spite of a claim\citep{lo09}.
Some authors showed that there exist close binary massive black holes 
with a short orbital period less than ten years and significant orbital eccentricity\citep{armi2,haya09,cu09}.

\citet{haya07} studied 
the accretion flow from a circumbinary disk onto binary massive black holes,
using a smoothed particle hydrodynamics (SPH) code.
They found that mass transfer occurs from the circumbinary disk to each black hole. 
The mass accretion rate significantly depends on the binary orbital phase in eccentric binaries, 
whereas it shows little variation with orbital phase in circular binaries.
Periodic behaviors of the mass accretion rate in the binary system 
with the different geometries or system parameters were also discussed 
by some other authors\citep{bog08,mm08,cu09}.
Recently, \citet{bo09} and \citet{do09} proposed the hypothesis 
that SDSSJO92712.65+294344.0 consists of two massive black holes in binary, 
by interpreting the observed emission line features as those arising from 
the mass-transfer stream from the circumbinary disk.

\citet{haya08} have, furthermore, performed a new set of simulations at higher resolution with an 
energy equation based on the blackbody assumption, 
adopting the same set of binary orbital parameters we had 
previously used (Hayasaki et al. 2007) ($a=0.01\,\rm{pc}$, eccentricity $e=0.5$, 
and mass ratio $q=1.0$). 
By this two-stage simulation, they found that 
while the Optical/NIR light curve exhibits little variation,
the X-ray/UV light curve shows significant orbital modulation
in the triple-disk system, which consists of an accretion disk around each black hole 
and a circumbinary disk around them. X-ray/UV periodic light variation 
are originated from a phase-dependent mass transfer from circumbinary disk (cf. \cite{khato05}).
The one-armed spiral wave on the accretion disk induced by 
the phase-dependent mass transfer causes the mass 
to accrete onto each black hole within one orbital period.
This is repeated every binary orbit. 
These unique light curves are, therefore, expected to be 
one of observational signatures of binary massive black holes.

Highly sensitive X-ray monitors over a wide area provide us with
a unique opportunity to discover close binary massive black holes in an
unbiased manner, based on the detection of the orbital flux
modulation. Monitor of All-sky X-ray Image (MAXI; \cite{ma09}), 
a Japanese experimental module attached to the International Space Station, 
is now successful in operation since the launch in 2009 July. MAXI, covering the energy band of 0.5--30 keV,
achieves a significantly improved sensitivity as an all X-ray monitor
compared with previous missions. According to the hard X-ray
luminosity function of AGNs by \citet{ueda03}, $\approx$1,300
nearby AGNs can be detected at the confusion flux limit of
$\sim$0.2~mCrab from the extragalactic sky at galactic latitudes
higher than 10$^\circ$. Among them, the brightest $\sim$100 AGNs can
be monitored every 2 months with a flux accuracy of 20\% level. Over
the plausible mission life of MAXI ($\ge2\,\rm{yr}$), it is possible to detect binary
massive black holes in nearby AGNs with binary orbital periods less
than $10\,\rm{yr}$. Besides MAXI, the Swift/Burst Alert Telescope (BAT) 
survey \citep{tu09} can make a similar job in the hard X-ray band of 15--200 keV.

In this paper, we investigate mass functions of binary massive black holes
with a very-short orbital period detectable with MAXI and/or Swift/BAT.
The plan of this paper is organized as follows. 
In Section~2, we describe the evolutionary scenario 
of binary massive black holes 
in the framework of coevolution of 
massive black holes and their host galaxies. 
Section~3 shows mass functions of 
close binary massive black holes.
They can be written as the product of the
observed black-hole mass function of nearby AGNs and probability 
for finding binary massive black holes, 
based on the evolutionary scenario as described in Section~2.
Brief discussions and conclusions are summarized in Section~4.


\section{Final-parsec evolution of binary massive black holes}
\label{2}
We first describe the evolution of binary massive black holes, 
focusing on interaction with surrounding gaseous disks 
in the framework of coevolution of massive black holes 
and their host galaxies.

Does the black hole mass correlate with the velocity dispersion 
of bulge in individual galaxies despite that there is a single or binary in their center? 
This is one of fundamental problems in the framework of the coevolution of massive black holes and their host galaxies. 
In some elliptical galaxies, there is a core with 
the outer steep brightness and inner shallow brightness. 
Binary massive black holes are considered to be closely associated 
with such a core structure with the mass of stellar light deficit\citep{ebi91,milo01}.
Recently, \citet{kb09} showed the tight correlations 
among black hole masses, velocity dispersions of host galaxies, 
and masses of stellar light deficits, 
using the observational data of 11 elliptical galaxies with cores.
This suggests that 
these correlations are still held for 
even if there are 
not only a single massive black hole but also binary massive black holes
in cores of elliptical galaxies.

Binary massive  black holes are considered mainly to evolve 
via three stages \citep{bege80}.
Firstly, each of black holes sinks independently 
towards the center of the common gravitational potential 
due to the dynamical friction with neighboring stars. 
If the binary can be regarded as a single black hole, 
its gravitational influence radius to the field stars is defined as
\begin{eqnarray}
r_{\rm{inf}}=\frac{GM_{\rm{bh}}}{\sigma_{*}^{2}}
\sim3.4[\rm{pc}]
\left(\frac{M_{\rm{bh}}}{10^7M_\odot}\right)^{1-2/\beta_2},
\label{eq:rinf}
\end{eqnarray}
where the tight correlation between the black hole mass
and one-dimensional velocity dispersion, $\sigma_*$, of the stars, 
the so-called $M_{\rm{bh}}-\sigma_*$ relation:
$M_{\rm{bh}}/10^7M_\odot=\beta_1\left(\sigma_*/200\rm{kms^{-1}}\right)^{\beta_{2}}$
is made use of. Unless otherwise noted, 
$\beta_1=16.6$ and $\beta_2=4.86$ are adopted by \citep{mm05} in what follows. 

When the separation between two black holes
becomes less than $1\,\rm{pc}$ or so, 
angular momentum loss by the dynamical friction 
slows down due to the loss-cone effect and a massive hard binary is formed.
This is the second stage.
The binary harden at the radius
where the kinetic energy per unit mass of the star 
with $\sigma_*$, equals to the binding energy per
unit mass of the binary\citep{q96}.
Its hardening radius is defined as
\begin{eqnarray}
a_{\rm{h}}=\frac{1}{4}\frac{q}{(1+q)^2}r_{\rm{inf}}
{
\sim8.5\times10^{-1}[\rm{pc}]
\frac{q}{(1+q)^2}
\left(\frac{M_{\rm{bh}}}{10^7M_\odot}\right)^{1-2/\beta_2},
}
\end{eqnarray}
where $q$ is the black-hole mass ratio.

Finally, 
the semi-major axis of the binary
decreases the radius at which the gravitational 
radiation dominates, and then a pair of black holes 
merge into a single supermassive black hole.
The detailed timescale in each evolutionary phase will be
 described in the following three subsections.
\subsection{{Star driven phase}}
\subsubsection{{The dynamical friction}}
\label{2.1}

Each black hole sinks into the common center of mass due to 
the dynamical friction with ambient field stars. 
The merger rate of two black holes is given by\citep{bt1987}
\begin{eqnarray}
\frac{\dot{a}(t)}{a(t)}=-\frac{0.428}{\sqrt{2}}\ln\Lambda\frac{GM_{\rm{bh}}}{\sigma_*}\frac{1}{a^2(t)},
\label{tdfrate}
\end{eqnarray}
where $a(t)$ is the separation between two black holes and $\ln\Lambda\approx10$ is the Coulomb logarithm.
The decaying timescale of black-hole orbits is then written
\begin{eqnarray}
&& t_{\rm{df}}=\biggr|\frac{a(t)}{\dot{a}(t)}\biggr|
{
\sim8.4\times10^{6}[\rm{yr}]
\left(\frac{a(t)}{a_{0}}\right)^2
\left(\frac{M_{\rm{bh}}}{10^7M_{\odot}}\right)^{1/\beta_2-1},
}
\label{tdf}
\end{eqnarray}
where $a_0=100\rm{pc}$ is the typical core radius of the host galaxy.
The integration of merger rate gives the following equation
\begin{eqnarray}
\frac{a(t)}{a_{0}}
=\left(1-\frac{t}{t_{\rm{c}}^{\rm{df}}}\right)^{1/2},
\label{at0}
\end{eqnarray}
where
\begin{eqnarray}
&&
t_{\rm{c}}^{\rm{df}}
\sim4.2\times10^{6}[\rm{yr}]
\left(\frac{a_{0}}{100\rm{pc}}\right)^{2}
\left(\frac{M_{\rm{bh}}}{10^7M_{\odot}}\right)^{1/\beta_2-1}.
\label{tcdf}
\end{eqnarray}
Recall that $\beta_2\sim5$, and hence the index of mass dependence is $-4/5$ for both $t_{\rm{df}}$ and $t_{\rm{c}}^{\rm{df}}$.
\subsubsection{{Stellar scattering}}

Even after hardening of  the binary, the binary orbit continues 
to decay by loss cone refilling of stars due to the two-body relaxation.
In addition, the repeated gravitational slingshot 
interactions with stars makes the orbital decay significantly rapid 
for black hole with mass less than few$\times10^{6}M_\odot$.
For the system with a singular isothermal sphere,
the timescale is given as\citep{milo03}
\begin{eqnarray}
t_{\rm{ss}}=\biggr|\frac{a(t)}{\dot{a}(t)}\biggr|
\sim3.0\times10^8 [{\rm{yr}}]
\left(\frac{a_{\rm{h}}}{a(t)}\right)\left(\frac{M_{\rm{bh}}}{10^7M_\odot}\right).
\end{eqnarray}
For the black hole with mass greater than $10^{6.5}M_\odot$,
this mechanism contributes inefficiently to the orbital decay
of the binary on the subparsec scale.
Instead, the binary-disk interaction is likely to be
a dominant mechanism of the orbital decay. 
Quite recent N-body simulations show that stellar dynamics alone can also
resolve the final parsec problem (e.g., \cite{be09}).

\subsection{Gaseous-disk driven phase}

The circumbinary disk would be formed inside the gravitational influence radius 
after hardening of the binary. The inner edge of circumbinary disk 
is then defined as 
\begin{eqnarray}
&&
r_{\rm{in}}=\left(\frac{m+1}{l}\right)^{2/3}a(t)
\nonumber \\
&&
\sim1.8[{\rm{pc}}]\frac{q}{(1+q)^2}\left(\frac{M_{\rm{bh}}}{10^7M_\odot}\right)^{1-2/\beta_2}
\left(\frac{a(t)}{a_{\rm{h}}}\right),
\label{eq:rinh}
\end{eqnarray}
where ${a(t)}$ is the semi-major axis of binary, and $m=2$ and $l=1$ are adopted unless otherwise noted\citep{al94}. 

For simplicity, the circumbinary disk is assumed to be 
a steady, axisymmetric, and geometrically thin 
with a differential rotation{
and fraction of Eddington accretion rate:
\begin{eqnarray}
&&
\dot{M}_{\rm{acc}}=\eta\dot{M}_{\rm{Edd}}
\nonumber \\
&&
\sim2.2\times10^{-2}\left[\frac{M_{\odot}}{\rm{yr}}\right]
\left(\frac{\eta}{0.1}\right)\left(\frac{0.1}{\epsilon}\right)\left(\frac{M_{\rm{bh}}}{10^{7}M_{\odot}}\right),
\label{eq:accrate}
\end{eqnarray}
where $\eta$, $\epsilon$ and $\dot{M}_{\rm{Edd}}$ are
the Eddington ratio, mass-to-energy conversion efficiency, and 
 Eddington accretion rate defined by $\dot{M}_{\rm{{E}dd}}=
(1/\epsilon)
4\pi GM_{\rm{bh}}m_{\rm{p}}
/c\sigma_{\rm{T}}
$, where $m_{\rm{p}}$, $c$, and $\sigma_{\rm{T}}$ show the proton mass, light velocity, 
and Thomson scattering cross section, respectively.

The surface density of circumbinary disk can be then written as
\begin{eqnarray}
\Sigma=\frac{\dot{M}_{\rm{acc}}}{2\pi{\nu}}\biggr|\frac{d\ln{r}}{d\ln\Omega}\biggr|.
\label{sigma}
\end{eqnarray}
where the eddy viscosity is defined as $\nu=\alpha{c_{\rm{cv}}}H$ with the 
Shakura-Sunyaev viscosity parameter\citep{ss73}, $\alpha$, 
characteristic velocity, $c_{\rm{cv}}$, and disk scale-hight, $H$.

The stability criterion for self-gravitation of the disk is defined by
the Toomre Q-value: $Q=\Omega{c_{\rm{cv}}}/\pi{G\Sigma}$. 
If the disk structure obeys a {\it standard disk} with $\dot{M}_{\rm{acc}}$ with $\eta=0.1$,  
$Q$ is much less than 1.
This means that the disk is massive enough to be gravitationally unstable.
Therefore, we introduce a self-regulated, self-gravitating disk model (\cite{mu96, be97}).
The condition of self-regulated disk is given by
\begin{eqnarray}
Q\approx1.
\label{src}
\end{eqnarray}
From equations (\ref{sigma}) and (\ref{src}), 
the effective sound velocity of self-gravitating disk is written as
\begin{eqnarray}
c_{\rm{sg}}=
\left[\frac{G\dot{M}_{\rm{acc}}}{2\alpha_{\rm{sg}}}\biggr|\frac{\ln{r}}{\ln\Omega}\biggr|\right]^{1/3},
\end{eqnarray}
where we adopt $\alpha=\alpha_{\rm{sg}}\lesssim0.06$ {and $c_{\rm{cv}}=c_{\rm{sg}}$.
If the radiative cooling in the disk is so efficient, the disk would fragment. 
The criterion whether the disk fragment or not is given by $\alpha_{\rm{sg}}=0.06$\citep{rice05}. 
If $\alpha_{\rm{sg}}>0.06$, the fragmentation occurs in the disk and 
causes the subsequent star formation. Such a situation is beyond the scope of our disk model.}
Assuming that the disk face radiates as a blackbody,  
the sound velocity can be written
\begin{eqnarray}
c_{\rm{s}}=\left(\frac{R_{\rm{g}}}{\mu}\right)^{1/2}
\left(\frac{3GM_{\rm{bh}}\dot{M}_{\rm{acc}}}{8\pi r^3\sigma}\right)^{1/8},
\end{eqnarray}
where $R_{\rm{g}}$, $\mu$, and $\sigma$ are the gas constant, molecular weight, and Stefan-Bolzmann constant,
respectively. 

The self-gravity of the disk is stronger than the gravity of the central black hole 
at the self-gravitating radius where the total disk mass equals to the black hole mass.
The rotation velocity of the disk then become flat outside the self-gravitating radius:
\begin{eqnarray}
&&
r_{\rm{sg}}
\approx\frac{GM_{\rm{bh}}}{8}\left(\frac{G\dot{M}_{\rm{acc}}}{2\alpha_{\rm{sg}}}\right)^{-2/3}
\nonumber \\
&&
{
\sim64[\rm{pc}]
\left(\frac{\alpha_{\rm{sg}}}{0.06}\right)^{2/3}
\left(\frac{0.1}{\epsilon}\right)^{-2/3}
\left(\frac{\eta}{0.1}\right)^{-2/3}
\left(\frac{M_{\rm{bh}}}{10^7M_\odot}\right)^{1/3},
}
\label{rsg}
\end{eqnarray}
where we put $c_{\rm{cv}}=c_{\rm{sg}}$.
Inside $r_{\rm{sg}}$, the angular frequency of the disk corresponds 
to Keplerian one where the gravity of central black hole dominates
the dynamics of the disk.
When $c_{\rm{cv}}=c_{\rm{s}}$, the disk transits 
from the self-regulated, self-gravitating disk to standard disk.
Its radius is given as
\begin{eqnarray}
&&
r_{\rm{stsg}}=\left(\frac{R_{\rm{g}}}{\mu}\right)^{4/3}
\left(\frac{G\dot{M}_{\rm{acc}}}{2\alpha_{\rm{sg}}}\right)^{-8/9}
\left(\frac{3GM_{\rm{bh}}\dot{M}_{\rm{acc}}}{8\pi\sigma}\right)^{1/3}
\nonumber \\
&&
{
\sim4.4\times10^{-4}
[\rm{pc}]
\left(\frac{\alpha_{\rm{sg}}}{0.06}\right)^{8/9}
\left(\frac{0.1}{\epsilon}\right)
\left(\frac{\eta}{0.1}\right)
\left(\frac{M_{\rm{bh}}}{10^7M_\odot}\right)^{-2/9}.
}
\label{rstsg}
\end{eqnarray}
As the binary evolves, the disk structure gets to depend on black hole mass. 
Since $r_{\rm{in}}(a_{\rm{h}})$ is less than $r_{\rm{sg}}$ and 
more than $r_{\rm{stsg}}$ in the all black-hole mass range, the circumbinary disk is initially modeled as 
the self-regulated, self-gravitating disk with the Keplerian rotation.
When the semi-major axis decays to the decoupling radius defined by equations~(\ref{adsq}) and (\ref{adss}), 
$r_{\rm{in}}(a_{\rm{d}})$ is less than $r_{\rm{stsg}}$ for $10^5M_\odot\le M_{\rm{bh}}\lesssim3\times10^{6}M_\odot$.
The disk structure for $10^5\le M_{\rm{bh}}\lesssim3\times10^{6}M_\odot$ can then be described by the standard disk theory,
whereas the disk still remains to be the self-regulated, self-gravitating disk with Keplerian rotation 
for other black-hole mass ranges (see the dotted line of Fig.~\ref{chavsm}). 

The circumbinary disk and binary 
exchanges the energy and angular momentum 
through the tidal/resonant interaction. 
For moderate orbital-eccentricity range, 
the torque of binary potential 
dominantly acts on the circumbinary disk 
at the 1:3 outer Lindblad resonance radius 
where the binary torque is balanced with the viscous torque\citep{al94}.

On the other hand, 
the circumbinary disk deforms to be elliptical by the tidal/resonant interaction. 
The density of gas is locally enhanced by the gravitational potential 
at the closest two points from each black hole 
on the inner edge of circumbinary disk \citep{haya07}.
The angular momentum of gas is removal by the locally enhanced viscosity, 
and thus the gas overflows from the two points to the central binary. 
An accretion disk is then formed around each black hole by the transferred gas\citep{haya08}. 
The mass transfer therefore adds its angular momentum to the binary
via two accretion disks\citep{haya09}.  

In a steady state, the mass transfer rate equals to the accretion rate, $\dot{M}_{\rm{acc}}$.
Since it is much smaller than the critical transfer rate defined by 
equation (41) of \citet{haya09}, the effect of torque 
by the mass transfer torque can be neglected.
The orbital-decay rate is then approximately written by equation (40) of \citet{haya09} as
\begin{eqnarray}
\frac{\dot{a}(t)}{a(t)}\approx
-\frac{\dot{J}_{\rm{cbd}}}{J_{\rm{b}}}
\sqrt{1-e^2},
\label{adot0}
\end{eqnarray}
where {$e$ is the orbital eccentricity of the binary and} the net torque, $\dot{J}_{\rm{cbd}}$, 
from the binary to circumbinary disk can be approximately written as
\begin{eqnarray}
\dot{J}_{\rm{cbd}}
\approx
{
3^{4/3}
}
\frac{(1+q)^2}{q}
\frac{1}{t_{\rm{vis,in}}}
\frac{M_{\rm{ld}}}{M_{\rm{bh}}}
\frac{J_{\rm{b}}}{\sqrt{1-e^2}},
\label{jdotcbd}
\end{eqnarray}
where {$t_{\rm{vis,in}}=r_{\rm{in}}^2/\nu_{\rm{in}}=\sqrt{GM_{\rm{bh}}r_{\rm{in}}}/(\alpha c_{\rm{cv}}^2)$} is the viscous timescale measured at the inner edge of disk 
and $M_{\rm{ld}}$ 
is the local disk mass defined as $M_{\rm{ld}}=\pi r_{\rm{in}}^2\Sigma_{\rm{in}}$.
From equation~(\ref{sigma}), the product of the viscous timescale and ratio of 
the black hole mass to local disk mass is given by 
\begin{eqnarray}
{
t_{\rm{vis,in}}
}
\left(\frac{M_{\rm{ld}}}{M_{\rm{bh}}}\right)^{-1}
=
2\frac{M_{\rm{bh}}}{\dot{M}_{\rm{acc}}}
\biggr|\frac{d\ln\Omega}{d\ln{r}}\biggr|.
\label{tvismld}
\end{eqnarray}
Substituting equation~(\ref{jdotcbd}) with equation~(\ref{tvismld}) 
into equation~(\ref{adot0}), the orbital-decay rate can be expressed by
\begin{eqnarray}
&&
\frac{\dot{a}(t)}{a(t)}
=-\frac{1}{t_{\rm{c}}^{\rm{gas}}},
\label{adot1}
\end{eqnarray}
where $t_{\rm{c}}^{\rm{gas}}$ is the characteristic timescale of 
orbital decay due to the binary-disk interaction:
\begin{eqnarray}
t_{\rm{c}}^{\rm{gas}}
&&
=
{
\frac{2}{3^{4/3}}
}
\frac{q}{(1+q)^2}\frac{M_{\rm{bh}}}{\dot{M}_{\rm{acc}}}\biggr|\frac{d\ln\Omega}{d\ln{r}}\biggr|
\nonumber \\
&&
\sim 3.1\times10^{8}[\rm{yr}]
\frac{q}{(1+q)^2}
\left(\frac{0.1}{\eta}\right)
\left(\frac{\epsilon}{0.1}\right),
\label{tc0}
\end{eqnarray}
where we adopt $\Omega=\Omega_{\rm{K}}$.
Note that $t_{\rm{c}}^{\rm{gas}}$ is independent of the black-hole mass, semi-major axis, and viscosity parameter, but dependent on the black-hole mass ratio, Eddington ratio, and mass-to-energy conversion efficiency. These arise from the assumptions that the disk is axisymmetric with a fraction of Eddington accretion rate and its angular momentum is outwardly transferred by the viscosity of Shakura-Sunyaev type.

Integrating equation (\ref{adot1}), we obtain
\begin{eqnarray}
\frac{a(t)}{a_{\rm{h}}}=\exp{\left(-\frac{t}{t_{\rm{c}}^{\rm{gas}}}\right)}.
\label{atc}
\end{eqnarray}

Following equations~(29) of \citet{haya09}, 
the orbital eccentricity increases with time 
in the present disk model.
The orbital eccentricity is, however, 
expected to saturate during the disk-driven phase, 
because the angular momentum of the binary 
is mainly transferred to the circumbinary disk 
when the binary is at the apastron.
The saturation value of orbital eccentricity becomes $e=0.57$.
This value is estimated by equating 
the angular frequency at the inner edge of circumbinary disk 
with the orbital frequency at the apastron.

\subsection{Gravitational-wave driven phase}
{
The merging rate by the emission of gravitational wave 
can be written by\citep{p64} as
\begin{eqnarray}
\frac{\dot{a}(t)}{a(t)}=-\frac{256 G^3 M_\mathrm{bh}^3}{5 c^5 a^4}
\frac{q}{(1+q)^2}\frac{f(e)}{(1-e^2)^{7/2}},
\label{mrate_gw}
\end{eqnarray}
where $f(e)=1+73e^2/24+37e^4/96$.}
The coalescent timescale is then given as
\begin{eqnarray}
&&
t_\mathrm{gw}=\biggr|\frac{a(t)}{\dot{a}(t)}\biggr|
=\frac{5}{32}
\left(\frac{{a(t)}}{r_{\rm{S}}}\right)^4
\frac{r_{\rm{S}}}{c}
\frac{(1+q)^2}{q}(1-e^2)^{7/2} ,
\label{eq:tgr}
\end{eqnarray}
where $r_{\rm{S}}=2GM_{\rm{bh}}/c^2$ is the Schwarzschild radius.

The semi-major axis decays to the transition radius, 
$a_{\rm{t}}$ where the emission of gravitational wave 
is more efficient than the binary-disk interaction. 
In other words, 
$t_{\rm{gw}}$ becomes shorter than $t_{\rm{c}}^{\rm{gas}}$ inside the transition radius.
Comparing equation~(\ref{tc0}) with equation~(\ref{eq:tgr}), the transition radius is defined by  
\begin{eqnarray}
\frac{a_{\rm{t}}}{r_{\rm{S}}}=
\left[\frac{32}{5}\frac{ct_{\rm{c}}^{\rm{gas}}}{r_{\rm{S}}}
\frac{q}{(1+q)^2}\frac{f(e)}{(1-e^2)^{7/2}}
\right]^{1/4}.
\label{at}
\end{eqnarray}

When the timescale of orbital decay by the emission of gravitational radiation
is shorter than the viscous timescale measured
at the inner edge, the circumbinary disk is decoupled with the binary.
The decoupling radius is then defined by
\begin{eqnarray}
\frac{a_{\rm{d}}}{r_{\rm{S}}}
=\left[\frac{32}{5}\frac{ct_{\rm{vis,S}}}{r_{\rm{S}}}
\frac{q}{(1+q)^2}\frac{f(e)}{(1-e^2)^{7/2}}
\right]^{4/11}
\label{adsq}
\end{eqnarray}
for the standard disk,
and
\begin{eqnarray}
\frac{a_{\rm{d}}}{r_{\rm{S}}}
=\left[\frac{32}{5}\frac{ct_{\rm{vis,S}}}{r_{\rm{S}}}
\frac{q}{(1+q)^2}\frac{f(e)}{(1-e^2)^{7/2}}
\right]^{2/7}
\label{adss}
\end{eqnarray}
for the self-regulated, self-gravitating disk.
Here $t_{\rm{vis,S}}$ is the viscous timescale 
measured {
at the inner edge of circumbinary disk 
when $a(t)=r_{\rm{S}}$.
}
Depending on the disk model, it can be written as
\begin{eqnarray}
t_{\rm{vis,S}}
{
\sim5.9\times10^{2}[\rm{yr}]
\left(\frac{0.1}{\alpha_{\rm{SS}}}\right)
\left(\frac{0.1}{\eta}\right)^{1/4}
\left(\frac{\epsilon}{0.1}\right)^{1/4}
\left(\frac{M_{\rm{bh}}}{10^7M_\odot}\right)^{5/4}
}
\label{tvissq}
\end{eqnarray}
for the standard disk, and
\begin{eqnarray}
t_{\rm{vis,S}}
{
\sim5.6\times10^{4}[\rm{yr}]
\left(\frac{0.06}{\alpha_{\rm{sg}}}\right)^{1/3}
\left(\frac{0.1}{\eta}\right)^{2/3}
\left(\frac{\epsilon}{0.1}\right)^{2/3}
\left(\frac{M_{\rm{bh}}}{10^7M_\odot}\right)^{1/3}
}
\label{tvisss}
\end{eqnarray}
for the self-regulated, self-gravitating disk, respectively.

Integrating equation~(\ref{mrate_gw}), we obtain  
\begin{eqnarray}
\frac{a(t)}{a_{\rm{t}}}=\left(1-\frac{t}{t_{\rm{c}}^{\rm{gw}}}\right)^{1/4},
\label{atgw}
\end{eqnarray}
where $t_{\rm{c}}^{\rm{gw}}$ can approximately be written as
\begin{eqnarray}
&&
t_{\rm{c}}^{\rm{gw}}\simeq3.8\times
{
10^{11}[{\rm{yr}}]
\left(\frac{a(t)}{a_{\rm{t}}}\right)^4
\left(\frac{M_{\rm{bh}}}{10^7M_\odot}\right)^{-3}
}
\frac{(1+q)^2}{q}.
\nonumber \\
&&
\times
(1-e_{0}^2)^{7/2},
\label{tcgw}
\end{eqnarray}
{where} $e_{0}=0.57$ is the initial orbital eccentricity at $a_{\rm{t}}$.

{\subsection{Observable period range for binary black holes}}

The resonant/tidal interaction causes the mass transfer 
from the circumbinary disk to 
accretion disk around each black hole{\citep{haya07}}. 
The ram pressure by mass transfer acts on the outer edge of accretion disk
and gives a one-armed oscillation on the disk (cf. \cite{khato05}). 
The one-armed wave propagates from the outer edge to the black hole, 
which allows gas to accrete onto the black hole within the orbital period.
This is repeated every binary orbit.
This mechanism therefore originates periodic light variations synchronized 
with the orbital period{\citep{haya08}}.

For the observational purpose,
$a_{10}$ is defined as the semi-major axis corresponding 
to feasible orbital period, $10\rm{yr}$, detectable with
MAXI and/or Swift/BAT by
\begin{eqnarray}
a_{10}
\sim4.9\times10^{-3}[\rm{pc}]
\left(\frac{M_{\rm{bh}}}{10^7M_\odot}\right)^{1/3}.
\label{amax}
\end{eqnarray}
Fig.~\ref{porbvsm} shows the mass dependence 
of each orbital period evaluated at $a_{\rm{t}}$, $a_{\rm{d}}$, and $a_{10}$
for equal-mass binary massive black holes.
The dashed line, dotted line and horizontal dash-dotted line show 
the orbital period at $a_{\rm{t}}$, $a_{\rm{d}}$, and $a_{10}$, 
respectively.
The area filled in with the solid line shows the existential region 
of binary black hole candidates with periodic light-curve signatures 
detectable with MAXI and/or Swift/BAT.

Fig.~\ref{chavsm} shows the mass dependence on each characteristic semi-major axis, 
$a_{\rm{h}}$, $a_{\rm{t}}$, $a_{\rm{d}}$, and $a_{10}$, for the equal-mass binary.
All of the semi-major axes get longer as black hole mass is more massive.
The decoupling radius is described by equation ~(\ref{adsq}) when $M_{\rm{bh}}\lesssim3\times10^6M_\odot$, whereas it is described by equation~(\ref{adss}) when $M_{\rm{bh}}\gtrsim3\times10^6M_\odot$.
Note that $a_{10}$ is on the track, where the binary evolves by the binary-disk interaction,
when $M_{\rm{bh}}{\lesssim}2\times10^7M_\odot$, whereas 
$a_{10}$ is on the track, where the binary evolves by the emission of gravitational wave, when $M_{\rm{bh}}{\gtrsim}2\times10^7M_\odot $.

Fig.~\ref{tsvsa} shows the orbital-decay timescale, $|a/\dot{a}|$, 
of binary massive black holes with $M_{\rm{bh}}=10^{7.5}M_\odot$ 
in panel (a) and corresponding elapsed time in panel (b).
The dashed line shows the timescale in the first evolutionary phase
where the orbit decays by the dynamical friction (hereafter, dynamical-friction driven phase). 
The solid line shows the timescale in the second evolutionary phase where the orbit 
decays by the binary-disk interaction (hereafter, disk-driven phase).
The dotted line shows the timescale in the final phase where the orbit decays 
by the dissipation due to the emission of the gravitational wave (hereafter, gravitational-wave driven phase).
The dash-dotted line shows the orbital-decay timescale of by stellar scattering 
(hereafter, stellar-scattering driven phase). 
We note that the orbital-decay timescale of stellar-scattering driven phase 
is too long for stellar scattering to be efficient mechanism in binary evolution. 
The binary therefore evolves toward coalescence 
via first dynamical-friction driven phase, second disk-driven 
phase, and final gravitational-wave driven phase.
The orbital-decay timescale of disk-driven phase is the longest among the other two.

Fig.~\ref{tsvsa2} shows the orbital-decay timescale of the binary 
with the same format as that of panel (a) of Fig.~\ref{tsvsa}, 
but for $M_{\rm{bh}}=10^6M_\odot$.
Note that stellar scattering is efficient after hardening of the binary,
as shown in the dash-dotted line of Fig.~\ref{tsvsa2}.

\begin{figure}[ht!]
\resizebox{\hsize}{!}{
\includegraphics{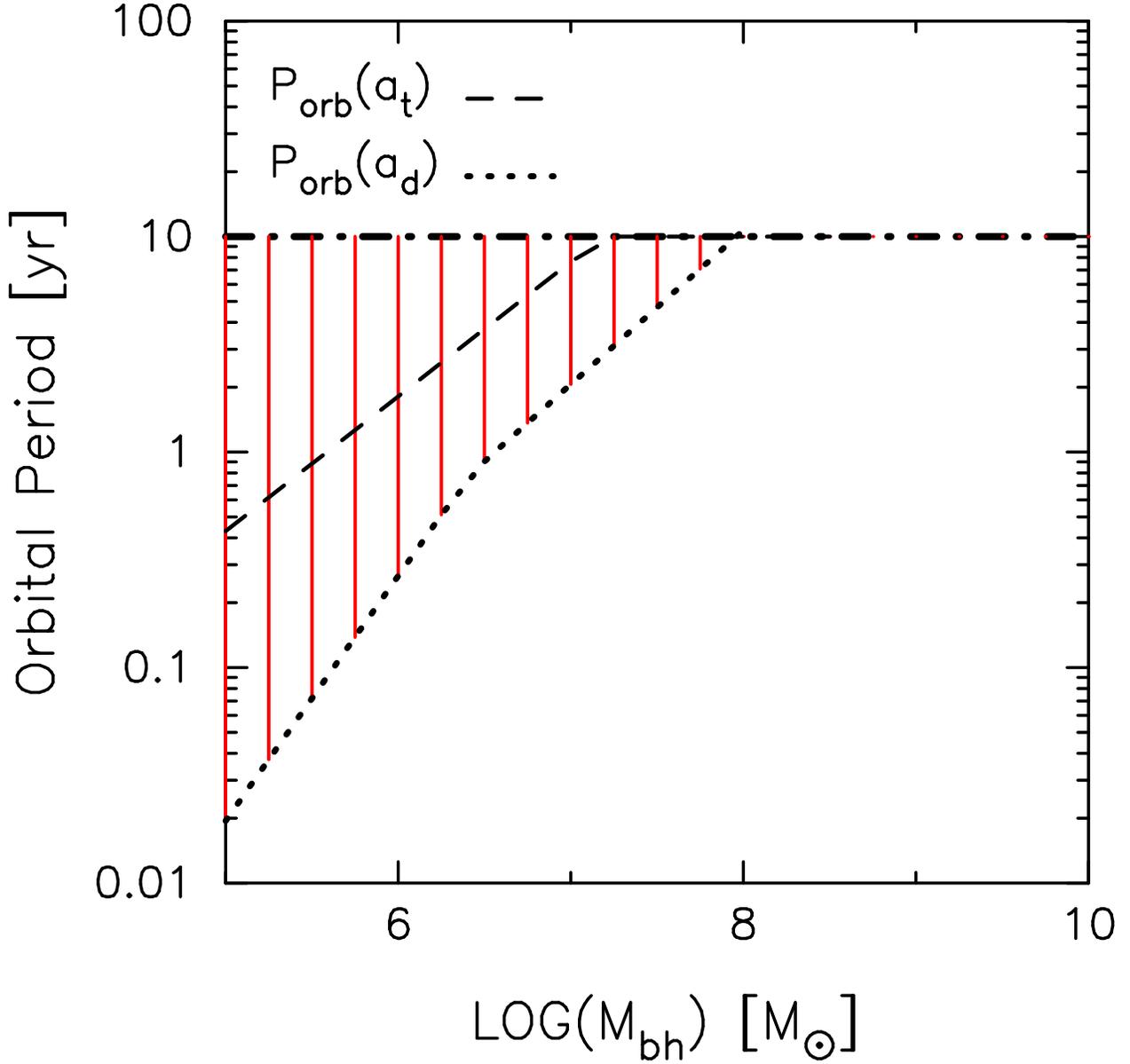}
}
  \caption{
Observable period range of binary massive black holes with characteristic semi-major axes. 
The dashed line shows the mass dependence of the orbital period evaluated at the transition radius, $a_{\rm{t}}$, where the dominant mechanism of binary evolution changes from the angular momentum loss by the binary-disk interaction to the dissipation by emitting gravitational wave radiation. The dotted line shows the one evaluated at the decoupling radius, $a_{\rm{d}}$, where the circumbinary disk decouples with binary massive black holes. The horizontal dash-dotted line shows the orbital period of ten years. The region drawn in solid lines shows the mass dependence of the orbital period less than ten years, which is detectable with MAXI and/or Swift.}
  \label{porbvsm}
\end{figure}

\begin{figure}[ht!]
\resizebox{\hsize}{!}{
\includegraphics{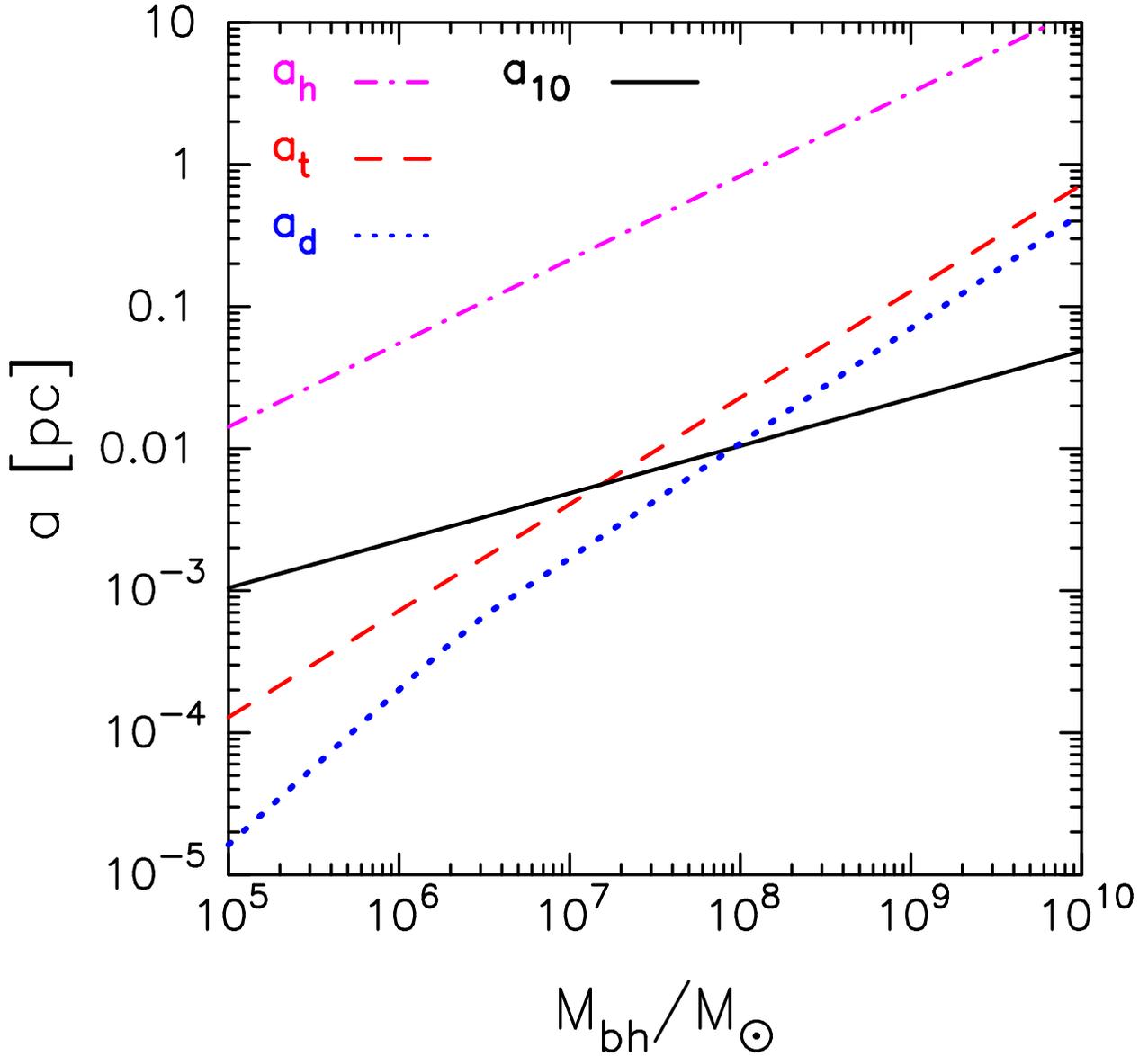}
}
  \caption{Mass dependence of characteristic semi-major axes of binary massive black holes. 
  The solid line shows the radius corresponding to the orbital period of ten years, $a_{10}$.
  The dashed line and dotted line show the transition radius, $a_{\rm{t}}$, and decoupling radius, $a_{\rm{d}}$, respectively. The dash-dotted line shows the hardening radius, $a_{\rm{h}}$, where the binding energy per unit mass of the binary equals to the kinetic energy of a star surrounding the binary. 
  }
  \label{chavsm}
\end{figure}

\begin{figure*}
\resizebox{\hsize}{!}{
\includegraphics*[width=86mm]{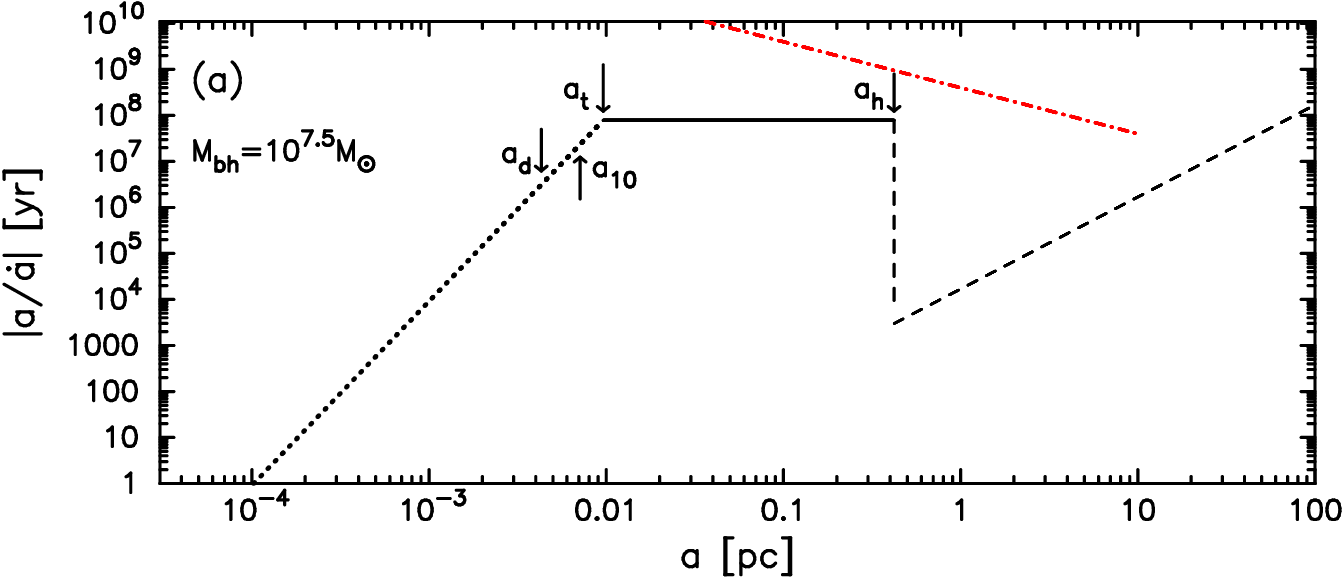}}
\resizebox{\hsize}{!}{
\includegraphics*[width=86mm]{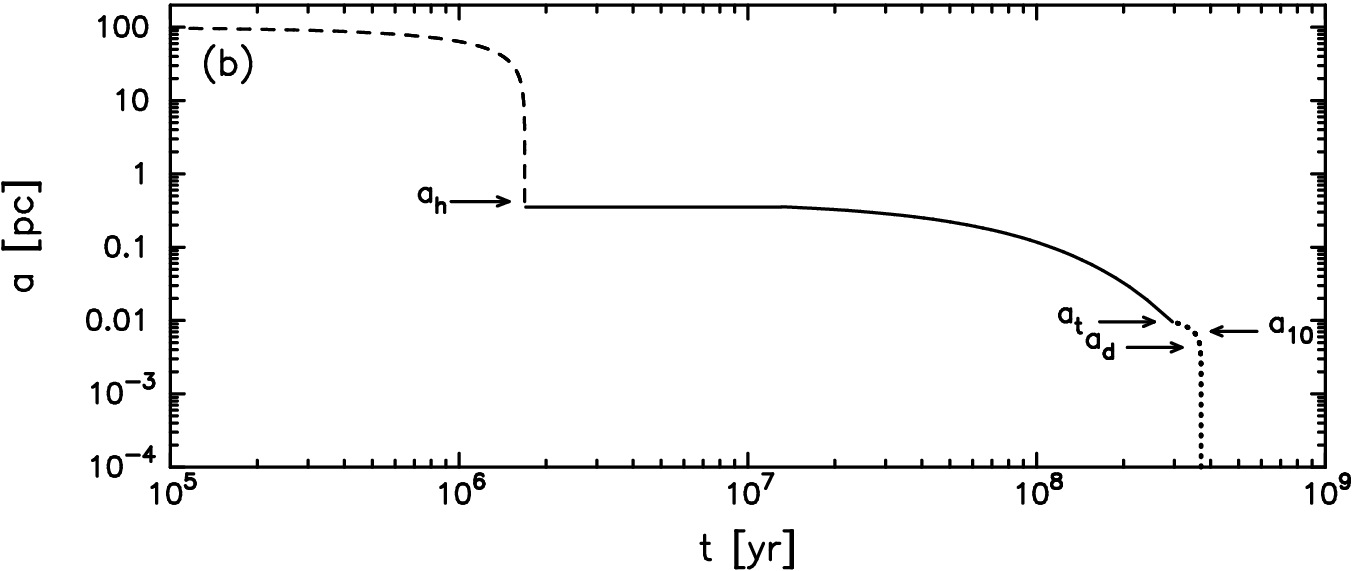}
}
\caption{
(a) Orbital-decay timescale, $|a/\dot{a}|$, of evolution of 
binary massive black holes with a semi-major axis from $100\rm{pc}$ to $10^{-4}\rm{pc}$. 
The total black hole mass is $M_{\rm{bh}}=10^{7.5}M_{\odot}$ with equal mass ratio, $q=1.0$.
(b) Corresponding elapsed time of evolution of binary massive black holes.
In both panels, the dashed line shows the first evolutionary phase 
in which two black holes get close each other toward the hardening radius, $a_{\rm{h}}$, 
by their angular momentum loss due to the dynamical friction with surrounding field stars.
The solid line shows the second evolutionary phase from $a_{\rm{h}}$ to $a_{\rm{t}}$
in which the angular momentum of the binary is removed 
by binary-disk interaction. The binary evolves from $a_{\rm{h}}$ to $a_{\rm{t}}$ 
where the stage changes from the disk-driven phase
 to the gravitational-wave driven phase in which binary massive black holes finally coalesce 
 by the emission of the gravitational radiation. 
There is the decoupling radius, $a_{\rm{d}}$, in the third evolutionary phase shown in the dotted line. 
In panel~(a), the dash-dotted line shows the timescale of orbital decay by stellar scattering.
}
\label{tsvsa}
\end{figure*}

\begin{figure*}
\resizebox{\hsize}{!}{
\includegraphics*[width=86mm]{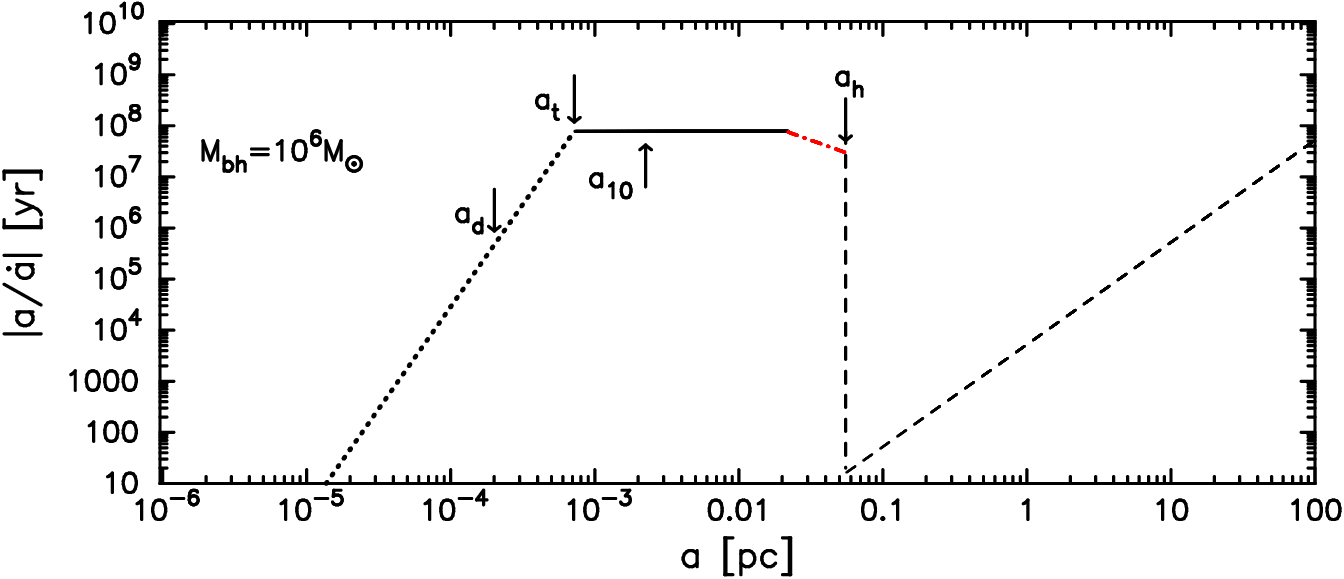}}
\caption{
Same format as panel~(a) of Fig.~\ref{tsvsa}, but for $M_{\rm{bh}}=10^6M_\odot$.
}
\label{tsvsa2}
\end{figure*}

\section{Mass function of binary massive black holes}

Observed black-hole mass function of nearby AGNs allows one to study
mass functions of binary massive black holes based on the evolutionary
scenario described in the previous section. Fig.~\ref{mfagn} shows the
fractional mass function of hard X-ray selected AGNs in the local
universe, where the black hole mass is estimated from the K-band
magnitudes of their host galaxies, as compiled by \citet{w09}. 
{This mass function is 99\% complete for the
uniform sample of local Seyfert galaxies, and hence the uncertainties
caused by the sample incompleteness should be regarded negligible.
}

Assuming that two galaxies
start randomly to merge during the interval, $t_{*}$, 
where $t_{*}$ is the look back time from present universe, 
a probability for finding binary massive black holes in the present universe
can be expressed by $|a/\dot{a}|/t_*$. 
The number of binary massive black holes, $N_{\rm{bbh}}$, 
in AGNs can be obtained by $N_{\rm{bbh}}=f_{\rm{c}}(|a/\dot{a}|/t_{*})/N_{\rm{AGN}}$, 
where $N_{\rm{AGN}}$ is the number of AGNs and $f_{\rm{c}}$ is 
a dimensionless fraction parameter.
More than $25\%$ of the host galaxies
of Swift/BAT AGNs show evidence for on-going mergers \citep{ko10}, 
where the separation between central two cores in the merging galaxies
is more than $kpc$ scale. 
Therefore, we set $f_{\rm{c}}=0.75$ unless otherwise noted, 
since we discuss binary black holes with a separation of much smaller scale.

One can estimate the probability for finding binary black holes
in nearby AGNs by putting $t_{\rm{AGN}}$ into $t_{*}$, 
where $t_{\rm{AGN}}=M_{\rm{bh}}/\dot{M}_{\rm{acc}}$ shows 
the e-holding accretion timescale which corresponds to the typical lifetime of AGNs. 
Fig.~\ref{probbbh} (a) and (b) show the mass-dependences of $|a/\dot{a}|/t_{\rm{AGN}}$, 
evaluated for $q=1.0$ and $q=0.1$, respectively.
The dashed line, dash-dotted line and dotted line show the probability 
for finding binary black holes with 
$a_{\rm{h}}$, $a_{\rm{t}}$, and $a_{\rm{d}}$, respectively.
It is noted from both panels that the probability 
of $a_{\rm{h}}$ is lower than that of 
$a_{\rm{t}}$ in the mass range less than the order of $10^6M_\odot$,
because the binary more rapidly evolves by stellar scattering than 
by disk-binary interaction as shown in the dash-dotted line of Fig.~\ref{tsvsa2}. 
The probability evaluated at $a_{\rm{t}}$ keeps constant over all of mass-ranges.

The solid line shows the integrated probability for finding binary black holes 
with the semi-major axis less than $a_{10}$. The integrated probability is approximately 
given by $\Delta{t}/t_{\rm{AGN}}$, where
\begin{eqnarray}
\Delta{t}=
\left\{
\begin{array}{ll}
t_{\rm{c}}^{\rm{gas}}
\ln\left(a_{10}/a_{\rm{t}}\right)
+t_{\rm{c}}^{\rm{gw}}(a_{t})
-
{
t_{\rm{c}}^{\rm{gw}}(a_{\rm{d}})
}
& a_{10}\ge a_{\rm{t}} \\
t_{\rm{c}}^{\rm{gw}}(a_{10})-t_{\rm{c}}^{\rm{gw}}(a_{\rm{d}})
& a_{10}\le a_{\rm{t}}, \\
\end{array}
\right.
\label{eq:5}
\end{eqnarray}
from equation~(\ref{tc0}), (\ref{atc}), (\ref{atgw}), and (\ref{tcgw}).

The integrated probability estimated for $a_{\rm{d}}\le{a}\le{a_{10}}$
is the monotonically decreasing function of black hole mass. 
Note that they rapidly decreases as the black hole mass becomes
greater than $10^{7}M_{\odot}$ for $q=1.0$ and $5\times10^7M_{\odot}$ for $q=0.1$.

Fig.~\ref{mfbbh} shows mass functions of binary massive black holes in AGNs. 
The mass function is defined by multiplying the black-hole mass function of AGNs
by the probability for finding binary black holes.
The mass functions evaluated at the hardening radius, $a_{\rm{h}}$ and 
transition radius, $a_{\rm{t}}$, are exhibited in panel~(a) and (b), respectively.
In both panels, the solid line and the dashed line show the mass function 
with $q=1.0$ and that with $q=0.1$, respectively.

There is less population of binaries with mass less than $10^{6.5}M_\odot$ in panel~(a), 
because the binary more rapidly evolves by stellar scattering than by disk-binary interaction 
in the mass range, as shown in the dash-dotted line of Fig.~\ref{tsvsa2}. 
Total fraction of binary black holes over all mass ranges are
$(4.3\pm2\%)$ for $q=0.1$ and $(13\pm4\%)$ for $q=1.0$ in both panels 
(the quoted errors reflect the statistical uncertainties in the Swift/BAT AGN mass function).
It is noted that from both panels that binary black holes of $M_{\rm{bh}}=10^{8.5-9}M_\odot$ 
are the most frequent in the nearby AGN population.

Fig.~\ref{mfbbh10} shows the mass functions of binary black holes 
with a constraint that the orbital period is less than ten years in both cases of $q=1.0$ and $q=0.1$.
From the figure, binary black holes of $M_{\rm{bh}}=10^{6.5-7}M_\odot$ for $q=1.0$ 
and those of $M_{\rm{bh}}=10^{7.5-8}M_\odot$ for $q=0.1$ are the most frequent in the nearby AGN population. 
It is notable that, assuming that all the binaries have equal black-hole mass ratio, 
$18\%$ of AGNs with black hole of $10^{6.5-7}M_{\odot}$ has binary black holes.

Total fraction of binary black holes over all mass ranges 
are $(1.8\pm0.6\%)$ for $q=1.0$ and $(1.6\pm0.4\%)$ for $q=0.1$.
We can therefore observe $15\sim27$ candidates for $1300$ AGNs detectable with MAXI, 
assuming that activities of all nearby AGNs lasts for $t_{\rm{AGN}}=M_{\rm{bh}}/\dot{M}_{\rm{acc}}$.
Note that MAXI covers the softer energy band (2--30 keV) than
Swift/BAT (15--200 keV), and hence the ratio of type-1 (unabsorbed)
AGNs to type-2 (absorbed) AGNs will be higher in the MAXI survey
($\approx$8:5 based on the model by \cite{ueda03}) than in the
Swift/BAT survey ($\approx$1:1, \cite{tu09}). Here we have
referred to the same AGN mass function, however, since we do not find
statistically significant difference between the observed mass
functions of type-1 and type-2 AGNs based on the Kolmogorov-Smirnov
test.
\section{Summary \& Discussion}

We study mass functions of binary massive black holes 
on the subparsec scale in AGNs based on the evolutionary 
scenario of binary massive black holes with surrounding gaseous disks 
in the framework of coevolution of massive black holes 
and their host galaxies.

As a very recent progress in observations of binary massive black holes 
with the Sloan Digital Sky Survey (SDSS), 
there is a claim that two broad emission line quasars 
with multiple redshift systems are subparsec binary candidates\citep{bl09}.
The temporal variations of such the emission lines are attributed to 
the binary orbital motion\citep{loeb09,sl09}.
These can be used as complementary approaches to search for 
binary massive black holes with MAXI and/or Swift/BAT.

Recently, \citet{vo09} predicted the fraction of binary quasars at $z<1$ based on the theoretical scenario for the hierarchical assembly of supermassive black holes in a $\Lambda$CDM cosmology. They adopted the merging timescale of binary black holes with a circumbinary disk estimated by \citet{ha09}, in order to explain the observed paucity of binary quasars in the SDSS sample (2 out of ~10000; \cite{bo09,do09,bl09}). For the black hole mass range of $\sim10^8M_\odot$, which these SDSS quasars likely have, our calculation gives a similar merging timescale ($\sim10^8$ year, independent of mass). Hence, our model will also be compatible with the SDSS results when applied to the same cosmological model. In a lower mass range, however, we predict a significantly longer merging timescale, by a factor of 10 at $\sim10^7M_\odot$, than that by \citet{ha09}, which rapidly decreases with the decreasing mass. Hence, much larger fractions of subparsec binary black holes are expected in our model than in \citet{vo09} if low mass black-holes are considered.

\citet{ks10} studied the nHz gravitational wave background generated by close binary massive black holes
with orbital periods between 0.1-10 years, taking account of both the cosmological merger rate and such the binary-disk interaction as the planetary (type II) migration\citep{ha09}. The orbital-decay timescale for low black-hole mass binaries ($M_{\rm{bh}}\le10^7M_\odot$) is much shorter than that of our model. This suggests that little stochastic gravitational wave background is attenuated by applying our model for their scenario, because the amplitude of 
gravitational wave background is proportional to the root of ratio of the orbital-decay timescale of the disk-driven phase to that of the gravitational-wave driven phase.

Our main conclusions are summarized as follows:
\begin{enumerate}
\renewcommand{\theenumi}{(\arabic{enumi})}
\item
Binary massive black holes on the subparsec scale 
can merge within a Hubble time by the interaction 
with triple disk consisting of an accretion disk around each black hole 
and a circumbinary disk surrounding them.
Assuming that the circumbinary disk is 
steady, axisymmetric, geometrically thin, self-regulated, self-gravitating 
but non-fragmenting with a fraction of Eddington accretion rate,
its orbital-decay timescale is given by $\sim3.1\times10^8q/(1+q)
(0.1/\eta)(\epsilon/0.1)[\rm{yr}]$, where 
$q$, $\eta$, and $\epsilon$ show the black-hole mass ratio, Eddington ratio, 
and mass-to-energy conversion efficiency, respectively.

\item
Binary black holes of $M_{\rm{bh}}=10^{8.5-9}M_\odot$ 
in the disk-driven phase 
are the most frequent among the AGN population.
Assuming that 
activities of all nearby AGNs lasts for the accretion timescale, 
$M_{\rm{bh}}/\dot{M}_{\rm{acc}}$, 
the total fraction of binaries with the semi-major axis 
evaluated at the hardening radius and transition radius
are estimated as $(4.3\pm2\%)$ and $(13\pm4\%)$, respectively.
\item
Assuming that all binary massive black holes have the equal mass ratio ($q=1.0$), 
$\sim20\%$ of AGNs with $M_{\rm{bh}}=
10^{6.5-7}M_\odot$ harbor binary black holes 
with orbital period less than ten years in their center.
This black-hole mass range therefore
provides the best chance to find such close 
binary black holes in AGNs.
\item
The total fraction of close binary massive black holes 
with orbital period less than ten years, 
as is detectable with MAXI and/or Swift/BAT, 
can be estimated as $(1.8\pm0.6\%)$ for $q=1.0$ and $(1.6\pm0.4\%)$ for $q=0.1$.
\end{enumerate}

\bigskip
We thank anonymous referee for the useful comments and suggestions.
We also thank Mike Koss and Richard Mushotzky for providing us with the
latest result on the merging rate of Swift/BAT AGNs before publication.
KH is grateful to Atsuo~T. Okazaki, Takahiro Tanaka, 
{and Stanley P. Owocki} for helpful discussions. 
This work has been supported in part by the Ministry of Education,
Science, Culture, and Sport and Technology (MEXT) through
Grant-in-Aid for Scientific Research (19740100, 18104003, 21540304, 22340045, 22540243 KH and 
20540230 YU, and 22740120 IN), and by the Grant-in-Aid for the Global COE Program ``The Next Generation of Physics, Spun from Universality and Emergence'' from MEXT of Japan.


\begin{figure}
\resizebox{\hsize}{!}{
\includegraphics{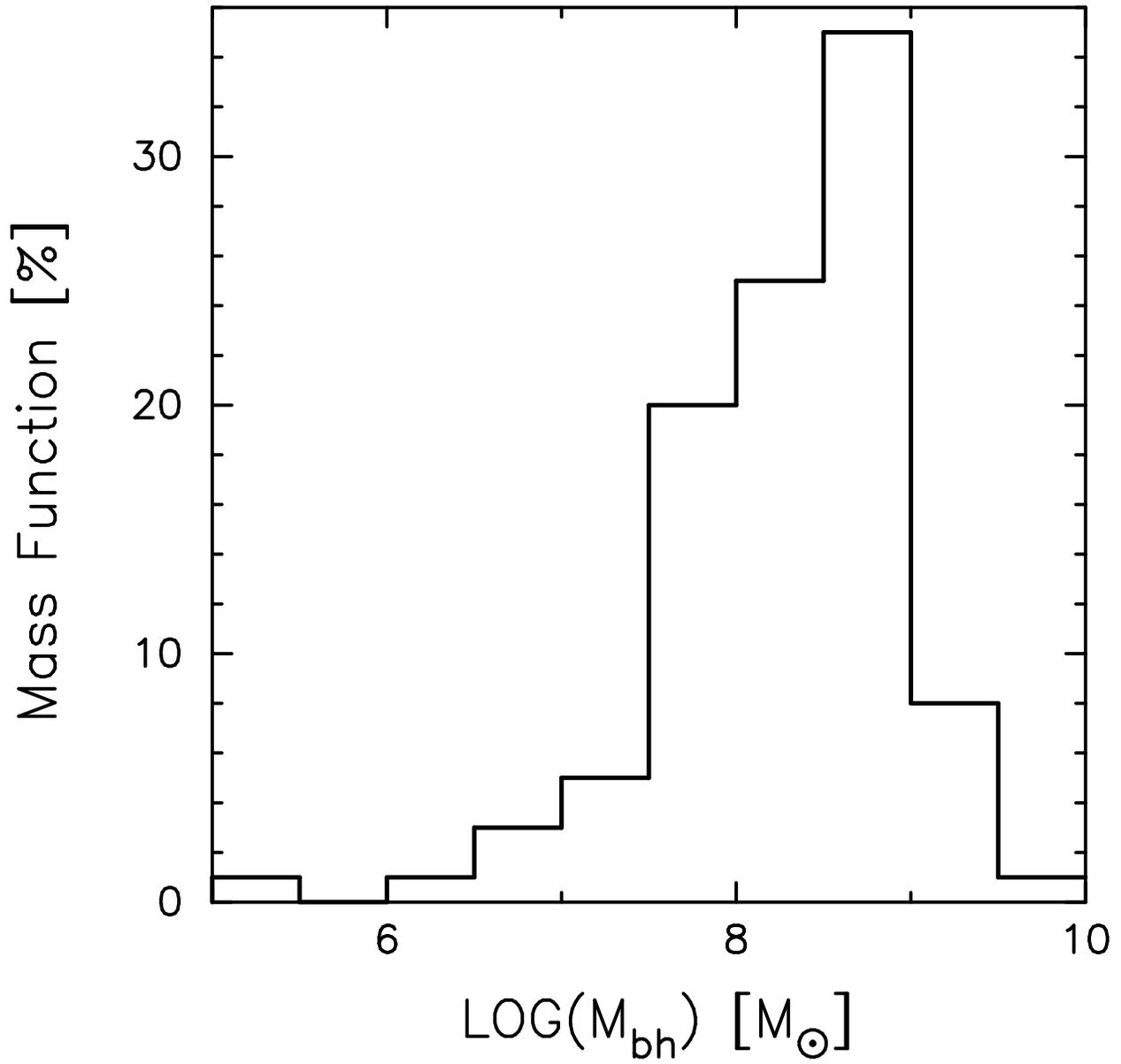}
}
  \caption{Normalized mass function of massive black holes in AGNs, 
detected with Swift/BAT (15--200 keV) given by \citet{w09}.}
  \label{mfagn}
\end{figure}

\begin{figure*}
\resizebox{\hsize}{!}{
\includegraphics{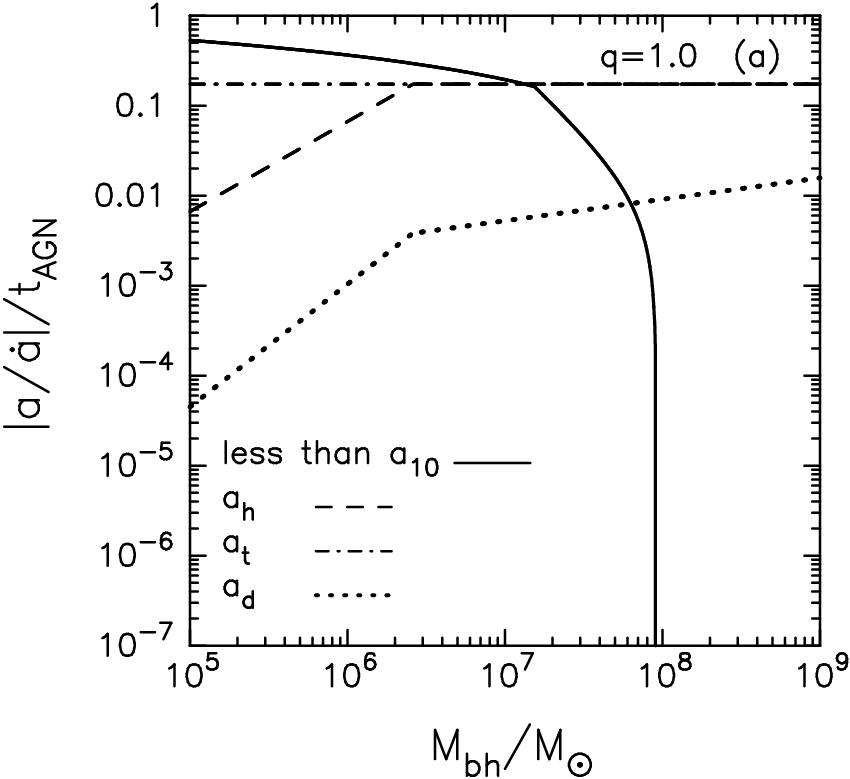}
\includegraphics{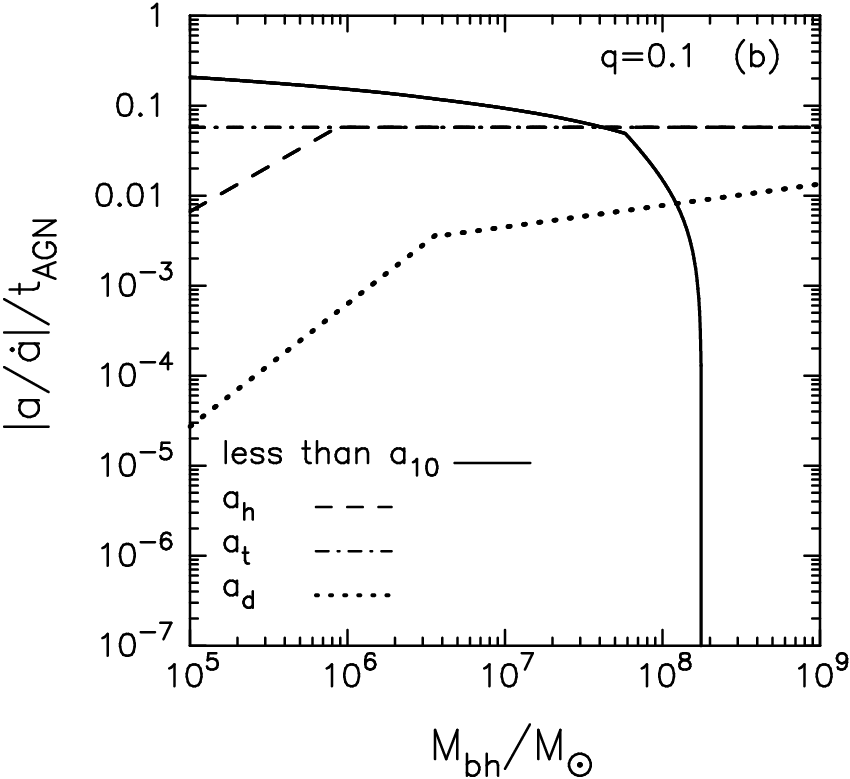}
}
  \caption{
  Mass dependence of probabilities for finding binary massive black holes evaluated for q=1.0 (a) and q=0.1 (b).
  The probability is defined as the orbital-decay timescale normalized by the accretion timescale, $|a/\dot{a}|/t_{\rm{AGN}}$, where $t_{\rm{AGN}}=M_{\rm{bh}}/\dot{M}_{\rm{acc}}$.
  The solid line shows the integrated probability for finding binary massive black holes 
  with the semi-major axis from $a_{10}$ to ${a_{\rm{d}}}$. 
  The dashed line, dash-dotted line, and dotted line show the probabilities evaluated 
  at $a_{\rm{h}}$, $a_{\rm{t}}$, and $a_{\rm{d}}$, respectively.
  }
  \label{probbbh}
\end{figure*}

\begin{figure*}
\resizebox{\hsize}{!}{
\includegraphics{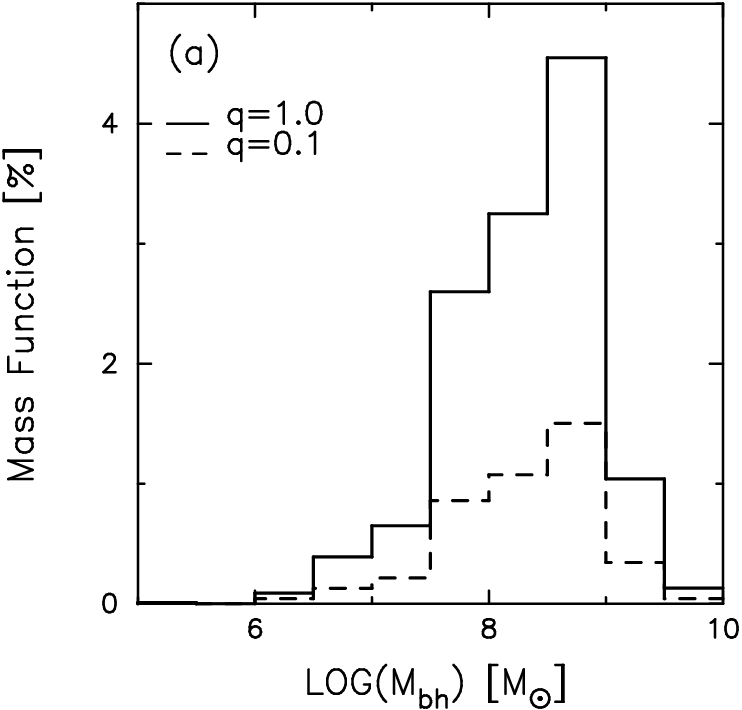}
\includegraphics{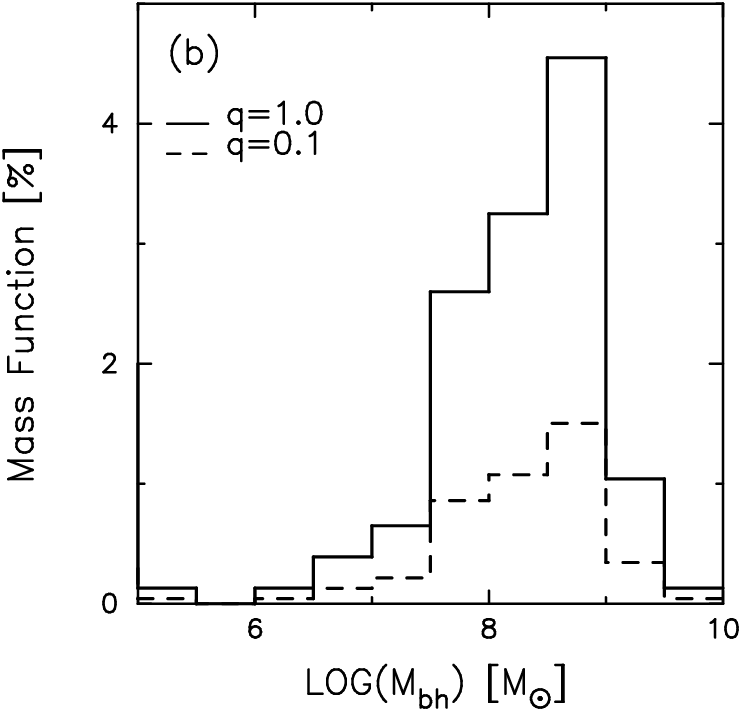}
}
  \caption{
  {
  Mass functions of binary massive black holes in AGNs 
  evaluated at the hardening radius, $a_{\rm{h}}$, (a) and transition radius, $a_{\rm{t}}$, (b). 
  They are defined by multiplying the mass function of AGNs 
  by the probability for finding binary massive black holes.
  }
  }
  \label{mfbbh}
\end{figure*}

\begin{figure}
\resizebox{\hsize}{!}{
\includegraphics{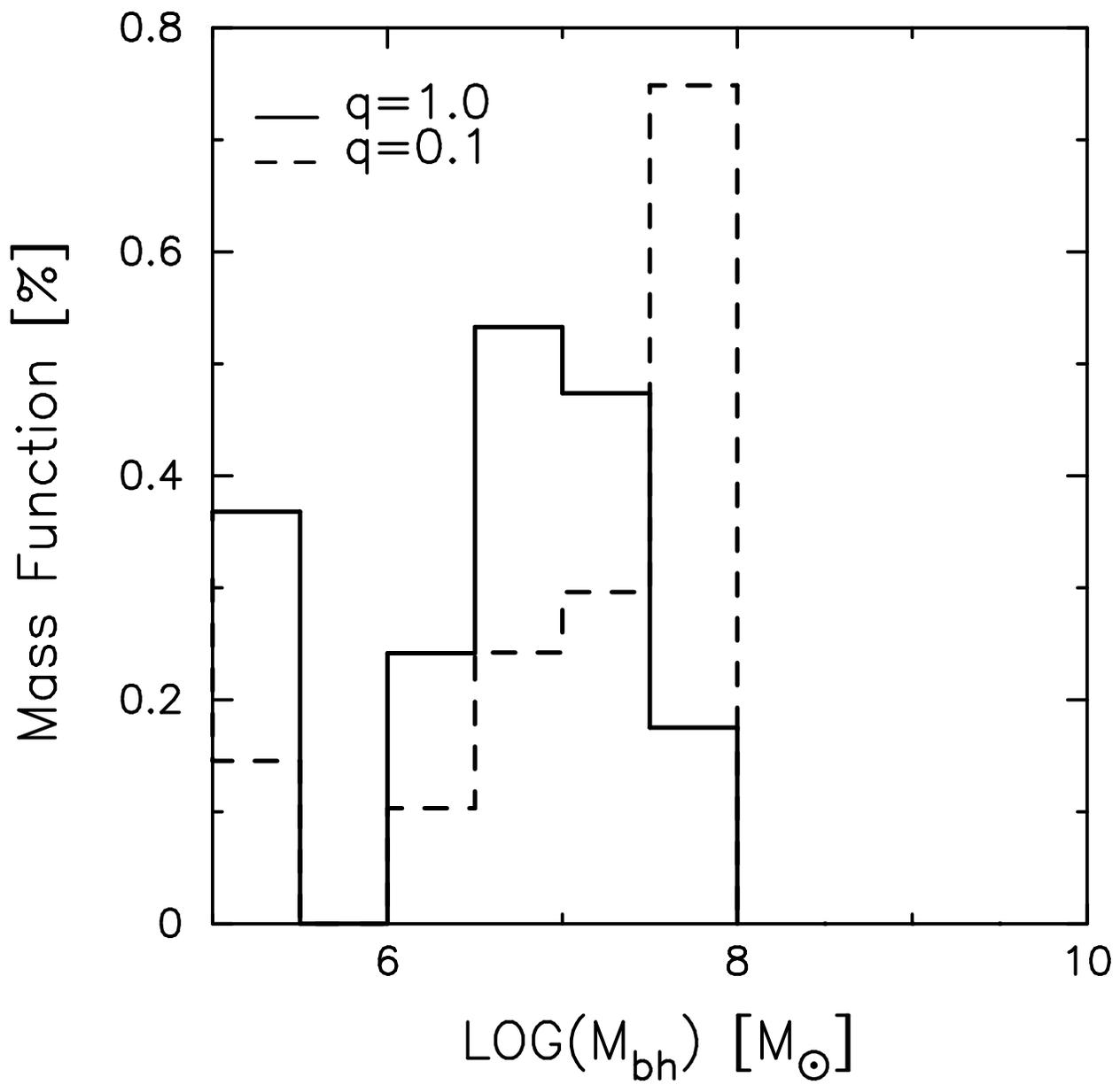}
}
  \caption{
  Mass functions of close binary massive black holes with the orbital period less than ten years.
  Same formats as Fig.~\ref{mfbbh}.
  }
  \label{mfbbh10}
\end{figure}


\end{document}